\documentclass[]{iopart}
\usepackage{iopams}

\usepackage{graphicx}
\usepackage{dcolumn}
\usepackage{bm}

\newcommand{\chiac}{$\chi_{\rm{ac}}$~}
\newcommand{\chiach}{$\chi_{\rm{ac}}(H)$~}
\newcommand{\chiact}{$\chi_{\rm{ac}}(T)$~}

\usepackage{color}

\begin{document}
\title[]{Confinement of Chiral Magnetic Modulations in the Precursor Region of FeGe}

\author{H. Wilhelm}
\address{Diamond Light Source Ltd., Chilton, Didcot, Oxfordshire, OX11 0DE, United Kingdom}

\author{M. Baenitz, M. Schmidt}
\address{Max Planck Institute for Chemical Physics of Solids,
N\"{o}thnitzer-Str. 40, 01187 Dresden, Germany}

\author{C. Naylor, R. Lortz}
\address{Department of Physics, The Hong Kong University of Science \& Technology, Clear Water Bay, Kowloon, Hong Kong}

\author{U. K. R\"{o}{\ss}ler, A. A. Leonov, A. N. Bogdanov}
\address{IFW Dresden, Postfach 270116, 01171 Dresden, Germany}

\begin{abstract}
We report on magnetic susceptibility and specific heat measurements
of the cubic helimagnet FeGe in external magnetic fields and temperatures
near the onset of long-range magnetic order at $T_C=278.2(3)$~K.
Pronounced anomalies in the field-dependent \chiach data as well as in the corresponding
imaginary part $\chi''_{\rm{ac}}(H)$ reveal a precursor region around $T_C$ in the magnetic phase diagram.
The occurrence of a maximum at $T_0=279.6$~K in the zero-field specific heat
data indicates a second order transition into a magnetically ordered state.
A shoulder evolves above this maximum as a magnetic field is applied.
The field dependence of both features coincides with crossover lines from the field-polarized
to the paramagnetic state deduced from $\chi_{\rm{ac}}(T)$ at constant magnetic fields.
The experimental findings are analyzed within the standard
Dzyaloshinskii theory for cubic helimagnets.
The remarkable multiplicity of modulated precursor states and the complexity
of the magnetic phase diagram near the magnetic ordering are explained by
the change of the character of solitonic inter-core interactions and
the onset of specific confined chiral modulations in this area.
\end{abstract}

\vspace{2pc}

\section{Introduction}\label{sec:introduction}
\begin{figure}
\begin{center}
\includegraphics[width= 0.9 \textwidth]{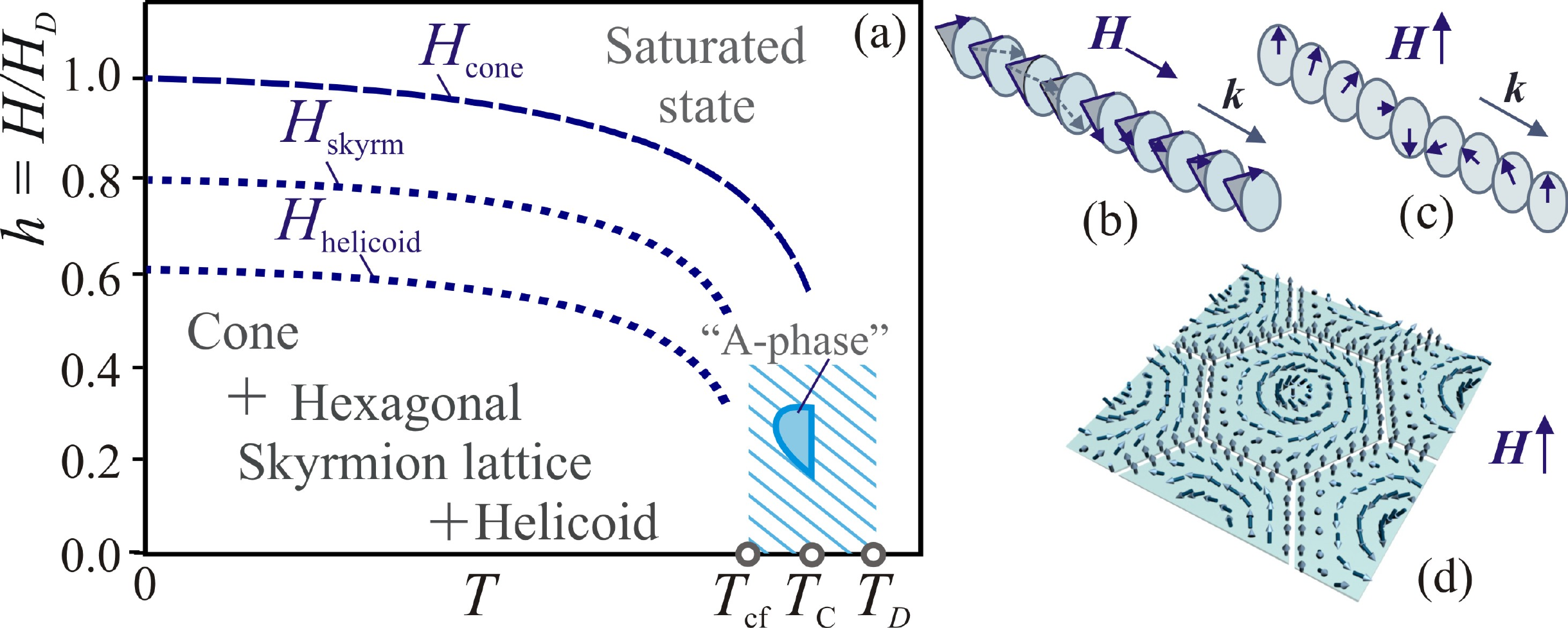}
\caption{%
(a) Schematic  magnetic phase diagram for cubic helimagnets based on experimental and theoretical findings.
Different magnetic states can exist and co-exist in applied magnetic fields $h=H/H_D$ ($H_D$ is a material specific field) up to certain
critical fields: conical helices, \textit{cones} (b), for fields $H<H_{\mathrm{cone}} (T)$ along the propagation direction $\bi{k}$,
\textit{helicoids} (c) as helical states distorted by a field perpendicular to $\bi{k}$ below $H_{\mathrm{helicoid}} (T)$,
and radial \textit{Skyrmions} which condense into a hexagonal lattice (d) in fields along the
Skyrmion axes for $H<H_{\mathrm{skyrm}} (T)$.
The small precursor region (hatched area) sets in above the \textit{confinement temperature} $T_{\mathrm{cf}}$ \cite{Roessler11,Leonov10}
and extends up to the \textit{nucleation} temperature $T_{\mathrm{D}}$ \cite{Roessler06}, where stable solitonic units can be nucleated.
The \emph{A} phase is located just below the Curie temperature $T_C$.
\label{phd1}}
\end{center}
\end{figure}

Cubic iron monogermanide, FeGe, \cite{Lebech89} together with the iso-structural MnSi \cite{Ishikawa76} and MnGe
\cite{Kanazawa11} as well as the pseudo-binary compounds Fe$_{1-x}$Co$_x$Si ($0.1\leq x\leq 0.7$) \cite{Beille81,Beille83,Grigoriev07,Grigoriev07b},
Mn$_{1-y}$Co$_y$Si ($0\leq y<0.9$) \cite{Beille83,Bauer10}, and  Mn$_{1-z}$Fe$_z$Si ($0\leq z<0.13$)
\cite{Bauer10,Grigoriev09} are non-centrosymmetric cubic helimagnets
(space group $P2_13$, $B20$-type structure)
with Curie temperatures $T_{\mathrm{C}}$ ranging from
a few Kelvin to almost ambient temperature.
The ground state of these magnets is well understood and consists of long-range helical modulations
with periods up to several hundred unit-cells (see table \ref{tab:parameter}).
The sense of the magnetization rotation is fixed due to the presence of the chiral Dzyaloshinskii-Moriya
(DM) interaction \cite{Dzyaloshinskii64,Bak80}.
However, in their magnetic field$-$temperature, ($H,T$), phase diagrams,
as sketched in figure \ref{phd1}(a), numerous puzzling physical anomalies
have been observed in a narrow temperature interval in the vicinity of
$T_{\mathrm{C}}$ \cite{Bauer10,Wilhelm11,Kusaka76,Wilkinson76,Komatsubara77,Kadowaki82,Ishikawa84,
Gregory92,Ishimoto95,Lebech95,Thessieu97,Pfleiderer04,Lamago06,Grigoriev06,Grigoriev06a,Stishov08,Petrova09,Pappas09,Muehlbauer09,
Neubauer09,Grigoriev10,Pappas11,Grigoriev11}.
The origin of these "precursor anomalies" (hatched area in figure \ref{phd1}(a)) and notably
the magnetic structure of the so-called \emph{A} phase is a long-standing and intriguing
problem in chiral magnetism.

\begin{table}
\caption{Curie temperature $T_C$, spontaneous magnetic moment $\mu_s (T\rightarrow 0)$,
lattice parameter $a$, helix period $L_D$, and saturation field $H_D\equiv H_{c2}(T=0)$
for some cubic helimagnets with \emph{B}20-type structure (space group $P2_13$).
 \label{tab:parameter}}
\begin{tabular}{llllllll}
\hline  
                        & $T_C$ (K)             & $\mu_s$ ($\mu_B$)    & $a$ (\AA)                & $L_D$ (\AA)           & $H_{D}$ (T)          \\ \hline
FeGe                    & 278.2 \cite{Wilhelm11}& 1 \cite{Lundgren68}  & 4.700 \cite{Richardson67}& 700 \cite{Lebech89}   & 0.359 \cite{Leonov12}   \\
MnSi                    & 29.5 \cite{Williams66}& 0.4 \cite{Williams66}& 4.559 \cite{Okada01}     & 180 \cite{Ishikawa76} & 0.62 \cite{Bloch75}    \\
MnGe$^{\rm{(a)}}$       & 170 \cite{Kanazawa11} & 1.7 \cite{Kanazawa11}& 4.795 \cite{Takizawa88}  & 30 \cite{Kanazawa11}  & 12 \cite{Kanazawa11}   \\
Mn$_{0.9}$Fe$_{0.1}$Si  & 6.8 \cite{Dyadkin10}  & $0.17^{\rm{(b)}}$    & 4.5529                   & 100 \cite{Grigoriev09}& 0.645 \cite{Grigoriev09}\\
Fe$_{0.65}$Co$_{0.35}$Si& 58.8  \cite{Beille81} & 0.218 \cite{Beille81}& 4.471 \cite{Shimizu90}   & 471 \cite{Beille83}   & 0.145 \cite{Beille81}   \\
\end{tabular}
\\$^{\rm{(a)}}$orders antiferromagnetically; $^{\rm{(b)}}$interpolated value using data in Ref~\cite{Bauer10}.
\end{table}

The theoretical investigations in the precursor region are closely connected
with the idea of chiral Skyrmions, i.e., axisymmetric solitonic states.
The prediction of a possible formation of such magnetic vortices in non-centrosymmetric ferromagnets in general
\cite{Bogdanov89,Bogdanov94} and in cubic helimagnets in particular \cite{Bogdanov05,Roessler06,Roessler11} have triggered
an intensive quest for these exotic states \cite{Grigoriev07b,Wilhelm11,Lamago06,Grigoriev06,Grigoriev06a,Pappas09,Muehlbauer09,
Neubauer09,Grigoriev10,Pappas11,Grigoriev11,Leonov10,Maleyev11}.
Of particular interest in the experimental investigations is the \textit{A} phase, a small area below $T_C$
reported for MnSi many years ago \cite{Komatsubara77,Kadowaki82}.
Within this pocket transverse chiral modulations attributed to helicoids have been identified \cite{Lebech95} (figure \ref{phd1}(c)) while outside this area a
one-dimensional \textit{cone} phase has been observed \cite{Ishikawa76} (figure \ref{phd1}(b)).
A possible magnetic state with transverse chiral modulation in applied field can also rely on skyrmionic textures
\cite{Bogdanov89,Bogdanov94,Roessler06,Roessler11,Leonov10} as a dense-packed hexagonal Skyrmion lattice
sketched in figure \ref{phd1}(d).
Subsequent intensive experimental investigations \cite{Bauer10,Lamago06,Grigoriev06,Grigoriev06a,Muehlbauer09,Neubauer09} led to conflicting conclusions about the magnetic state of the \emph{A} phase \cite{Wilhelm11,Pappas09,Grigoriev10,Pappas11}.
%
%
%
The existing experimental data are still inconclusive to reconstruct detailed  magnetic structures of the modulated
states arising in the precursor area and/or to prove the existence or non-existence of skyrmionic
states either in the \emph{A} pockets or in any other part of the precursor region.

Chiral Skyrmions have been finally discovered in thin layers
of FeGe \cite{Yu11} in a broad temperature range \textit{far below} the precursor region found in bulk samples.
Similar skyrmionic states have been observed in thin films of Fe$_{0.5}$Co$_{0.5}$Si \cite{Yu10},
MnSi \cite{TokuraTalks}, and in the non-centrosymmetric
cubic ferrimagnet Cu$_{2}$OSeO$_{3}$ \cite{TokuraTalks}.
These break-through investigations are based on the direct observation
of helical and skyrmionic states by Lorentz TEM.
They confirm the existence of essential features of chiral Skyrmions in non-centrosymmetric
magnets as proper static and  multidimensional topological solitons and the field-driven
formation of Skyrmion lattices, as theoretically anticipated
\cite{Bogdanov89,Bogdanov94,Bogdanov05,Roessler11,Butenko10}.

In contrast, the recent attempts to resolve and to understand the magnetic structures
in the precursor region and the temperature-driven magnetic ordering in \emph{bulk} chiral magnets
do not provide a clear picture \cite{Grigoriev07b,Bauer10,Wilhelm11,Lamago06,Grigoriev06,Grigoriev06a,Stishov08,
Petrova09,Pappas09,Muehlbauer09,Neubauer09,Pappas11}.
Nevertheless, these studies have accumulated an impressive database on peculiarities
of magnetic properties in this enigmatic area.
Together with earlier observations \cite{Kusaka76,Komatsubara77,Kadowaki82,Ishikawa84,Gregory92,Lebech95,Thessieu97}
and recent theoretical findings \cite{Wilhelm11,Roessler11,Leonov10}
the following important conclusions about the physical nature of the precursor region can be drawn:
(i) It consists of a  \textit{multiplicity} of modulated states  which have complex \textit{multi-dimensional} textures, and
(ii) precursor states are \textit{fundamentally} different from regular chiral modulations, as theoretically motivated
in \cite{Wilhelm11,Leonov10}.

The theoretical justification for the complexity of the precursor phenomena is based on the analysis of
rigorous solutions for modulated states from the basic phenomenological model of
chiral magnets near the ordering temperature \cite{Wilhelm11,Roessler11,Leonov10}.
It was found that the magnetic properties of chiral modulations strongly depend on the interactions
of their solitonic constituents, i.e., axisymmetric (\textit{double-twisted}) cell cores
in Skyrmion lattices and  \textit{kinks} in helicoids.
In the main part of the magnetic phase diagram chiral textures
consist of assemblies of localized particle-like modulations with \textit{repulsive} core-core interactions.
Near the Curie temperature, however, these interactions become \textit{attractive} as a result of the "softening"
of the magnetization amplitude \cite{Leonov10}.
These attractive inter-core interactions drastically influence the magnetic structures and properties
of the chiral modulations and are responsible for numerous anomalies in the precursor region
\cite{Wilhelm11,Roessler11,Leonov10}.

The outline of this article is as following:
Section \ref{sec:basicTheory} starts from a brief theoretical overview of chiral phases in cubic helimagnets including results for low-temperature modulations
\cite{Dzyaloshinskii64,Bak80,Bogdanov94,Bogdanov05} and rigorous solutions for helical and skyrmionic states near the ordering temperature \cite{Wilhelm11,Roessler06,Roessler11,Leonov10}.
Based on this, we reconstruct a schematic magnetic phase diagram of a cubic helimagnet (figure~\ref{phd1}) and expound a new picture of chiral precursor states introduced in \cite{Wilhelm11,Roessler11,Leonov10}.
Then we present in section \ref{sec:results} magnetic ac-susceptibility and specific heat measurements on FeGe in external fields in
the vicinity of the transition into the long-range ordered helical state at $T_C=278.2$~K \cite{Wilhelm11}.
FeGe provides a generic example of a chiral cubic helimagnet with a complex magnetic phase diagram in a very narrow precursor region around $T_C$ where some kind of magnetic fluctuations are perceived.
In section~\ref{sec:discussion} we discuss the relation between the observed results and theoretical solutions
for precursor states and compare the experimental findings with the calculated
phase diagram for a basic model within the framework of the phenomenological Dzyaloshinskii theory of cubic helimagnets.

\section{Magnetic phase diagram of cubic helimagnets}\label{sec:basicTheory}
The theoretical description of chiral cubic helimagnets can be
based on the phenomenological Dzyaloshinskii theory \cite{Dzyaloshinskii64}.
The magnetic states are found as energy minima of a (free) energy functional for the components of the
magnetization vector $\bi{M}$ \cite{Bak80}:
\begin{equation}
w=\underbrace{A  \left(\mathrm{grad}\bi{M} \right)^2
-D\,\bi{M}\cdot \mathrm{rot}\bi{M}
-\bi{M}\cdot\bi{H}}_{w_0(\bi{M})} + w_a (\bi{M}) + f(M)\,.
\label{density}
\end{equation}
This model includes the exchange stiffness with constant $A$
and the Dzyaloshinskii-Moriya (DM) interaction with constant $D$
as main magnetic couplings.
Together with the Zeeman energy they are the isotropic interactions $w_0(\bi{M})$.
The free energy also contains the much weaker anisotropy energy $w_a (\bi{M}) = \sum_{i=1}^{3}[K
\left(\partial M_i / \partial x_i \right)^2 +K_c M_i^4 ]$ with the exchange anisotropy $K$ and the
cubic magnetocrystalline anisotropy $K_c$ ($x_i$ are the components of the spatial variable).
Finally, the homogeneous magnetic free energy contribution $f(M)$ describes the variation of the magnetization
modulus $M \equiv |\bi{M}|$.
The \textit{basic} chiral modulations of the magnetic states are described by
the "isotropic" model, i.e., $w=w_0 (\bi{M})$. 
In this case the magnetization $\bi{M}$ has essentially a fixed length in a broad temperature range
and nonuniform magnetic states consist only of rotations or twists.

Let us recall the basic solutions for this model
\cite{Dzyaloshinskii64,Bak80,Bogdanov94,Bogdanov05}
which are essential for the discussions of the experimental results presented below.
The minimization of $w_0 (\mathbf{M})$ results in the phase diagram sketched in figure \ref{phd1}(a) and yields the following one-dimensional \cite{Dzyaloshinskii64,Bak80} and two-dimensional \cite{Bogdanov94,Bogdanov05}
chiral modulations depicted in figures \ref{phd1}(b-d):
(i) The \textit{cone phase}: It is a single harmonic mode propagating along
the field direction with period $L_D$ and flips into the saturated state at a critical
field $H_{\mathrm{cone}}(T)$ (figure \ref{phd1}(a)).
The saturation field of the cone phase at zero temperature $H_D \equiv H_{\mathrm{cone}} (0)\equiv H_{c2}(0)$
and the modulation period $L_D$ are given by
\begin{equation}
H_D = D^2M/(2A), \quad  L_D = 4 \pi A/|D|\,.
\label{cone}
\end{equation}
They provide characteristic, material specific parameters of the cubic helimagnets and are given for
several compounds in table \ref{tab:parameter}.
(ii) The \textit{helicoid}: It is a distorted helix with the propagation direction $\bi{k}$
perpendicular to the applied field (c.f. figure \ref{phd1}(c)).
(iii) The \textit{hexagonal Skyrmion lattice}:  This is a regular solution of the isotropic model
with "double-twist" modulations in a certain plane (figure \ref{phd1}(d)) and a homogeneous extension
along the third spatial direction \cite{Bogdanov94,Roessler06,Roessler11}.
At certain critical fields $H_{\mathrm{helicoid}} (T)$ and $H_{\mathrm{skyrm}}(T)$,
the helicoids and Skyrmion lattices transform into a set of localized structures (solitons), either
into isolated 360$^\circ$ domain walls (\textit{kinks}) or into isolated Skyrmions
by infinite expansion of the modulation period \cite{Dzyaloshinskii64,Bogdanov94}.
In particular, $H_{\mathrm{helicoid}} (0)= (\pi^2/16)H_D = 0.617 H_D$  \cite{Dzyaloshinskii64}
and $H_{\mathrm{skyrm}} (0) = 0.801 H_D$ \cite{Bogdanov94}.
By this transformation, the helicoids reveal themselves as kink-crystals, i.e.,
a succession of solitonic kinks, 360$^\circ$ domain wall sections with fixed shape and size of
their cores.
In the magnetic phase diagram (figure \ref{phd1}(a)) the critical fields
$H_{\mathrm{helicoid}} (T)$, $H_{\mathrm{skyrm}} (T)$, and $H_{\mathrm{cone}} (T)$
set the existence limit for the corresponding modulated states.

The free energy $f(M)$ becomes an important contribution to the isotropic model $w=w_0$ near the ordering temperature.
It is usually written in the canonic form of the Landau theory \cite{Bak80},
$f(M) = \alpha(T_{\mathrm{C}}^0 - T)M^2 +b M^4$,
where $T_{\mathrm{C}}^0$ is the Curie temperature of the bare ferromagnet.
However, the onset of magnetic order in a chiral magnet is shifted to a higher temperature
owing to the chiral DM interaction and the exchange shift $\Delta_D$.
The latter is given by material specific constants and determines the temperature width of
the precursor region.
Finally, we introduce a \textit{confinement} temperature $T_{\mathrm{cf}}$ that separates the phase diagram into two parts with different character of the interactions between chiral Skyrmions or kink cores (see figure \ref{phd1}(a)) \cite{Wilhelm11,Roessler11}.
All these parameters are given by
\begin{equation}
T_{\mathrm{C}} = T_{\mathrm{C}}^0 + \Delta_D/4, \quad
T_{\mathrm{cf}} = T_{\mathrm{C}} - \Delta_D, \quad
\Delta_D = D^2/(2 \alpha A) \ll T_{\mathrm{C}}^0\,.
\label{exchangeshift}
\end{equation}
The \emph{nucleation} temperature $T_{\mathrm{D}} = T_{\mathrm{C}} + \Delta_D/4 $
signifies the onset of stable double-twist excitations in the paramagnetic phase \cite{Roessler06}.

It is important to emphasize a fundamental distinction between a magnetic field-induced saturation of the cone phase
and the transitions of helicoids and Skyrmion lattices into the (field-polarized) homogeneous state.
%
The spin-flip-like saturation of the conical helix is a typical Landau \textit{symmetry-breaking} second order transition.
Here, upon decreasing field, the conical modulation sets in as the saturated state experiences an instability with respect to a particular mode.
The amplitude of the conical helix modulation with the propagation vector $\bi{k}$
is the small parameter of this \textit{instability-type} of transition \cite{DeGennes}.
At the transition, only this instable mode with a leading harmonic, expressed
by the Fourier coefficient $M_k$, is relevant even in the presence of anisotropies.
Approaching the transition from the ordered state, higher
harmonics $M_{nk}$ become unimportant, i.e., the ratio of Fourier coefficients $M_{nk}/M_{k} \rightarrow 0$.

The transition into the saturated state by setting free particle-like solitonic
states as 2D isolated Skyrmions and 1D kinks has an entirely different character.
These entities already exist in the saturated state as excitations with
a stable shape and size of their cores.
Therefore, this transition is a \textit{nucleation-type} of phase transition.
Thus, approaching such a nucleation transition from the ordered state of
a soliton lattice, all higher harmonic modes of a crystalline array
are retained, $M_{nk}/M_{k} \rightarrow \textrm{const}$, up to the transition point
where the lattice disintegrates into a gaseous set of free solitons.
As the Dzyaloshinskii theory of chiral magnets contains kinks and Skyrmions as
different and competing solitons, modulated states - with the exception of the simple
conical helix - are generically expected to form via the
nucleation-type of transition.

The character of inter-solitonic interactions in the saturated state ($H>H_{\mathrm{cone}}$), however, strongly conditions how exactly such a nucleation transition takes place.
As the formation of modulated states through nucleation of mesoscale structures always involves assembling stable solitonic entities, the nucleation transition in cubic helimagnets is expected to display similar mesophases as observed in the condensation or (partial) crystallization of liquid crystals or polymers \cite{Bouligand72}.

\section{Experimental Details}
For the present investigations single crystals of the cubic modification of FeGe, $\epsilon$-FeGe, were used.
Details about crystal growth and structural data are reported elsewhere \cite{Wilhelm07}.
The samples were checked with state-of-the-art characterization methods.
Subsequent thermodynamic, spectroscopic, x-ray and neutron measurements revealed
an excellent quality of the single crystals.
In the case of the ac-susceptibility measurements the crystal ($m=0.995$~mg) was aligned with the $[100]$-axis parallel to the external magnetic field in a Quantum Design PPMS.
Field sweeps were done in zero-field mode, i.e., the measuring temperature was always approached from 300~K in zero field.
The excitation field was 10~Oe and the measuring frequency was 1~kHz.
All ac-susceptibility data shown in the following represent the modulus $\chi_{\rm{ac}}=\sqrt{\chi'^2 +\chi''^2}$, with $\chi'$  and $\chi''$ the real and imaginary part of the susceptibility, respectively.
Given the almost spherical shape of the sample demagnetization effects are very small and can be neglected.
The specific heat was measured in a DC Peltier heat-flow calorimeter with continuous heating \cite{Plackowski02}.
It provides high relative resolution at a very high density of data points ($\Delta T = 0.05$~K) with a precision of the temperature measurement of about 0.1~K in the investigated temperature range.
The absolute value of the data has been calibrated through comparison with data measured in a commercial Quantum Design PPMS calorimeter.
The sample orientation with respect to the field was arbitrary.
The sweep rate of the temperature was kept at 1 K/min.
Up to 15 points/K were recorded and the data have been smoothed over 5 to 10 points.

\begin{figure}
\includegraphics[width=1\textwidth]{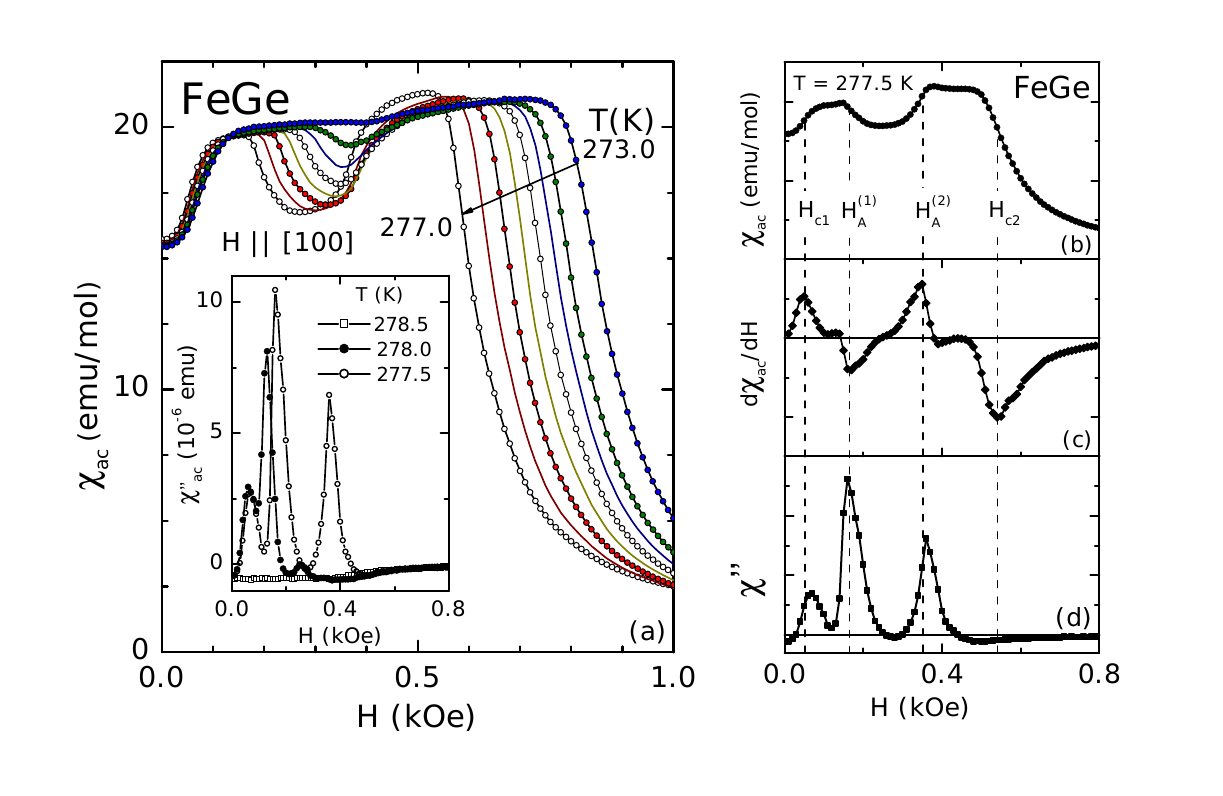}
\vspace*{-10mm}\caption{(a) Field dependence of \chiac for selected temperatures across the $A$ phase of FeGe.
The data sets between the two indicated temperatures were taken at $T=$ 276.5, 276, 275.5, 275, and 274~K.
Inset: The imaginary part of the susceptibility, $\chi''_{\rm{ac}}(H)$, shows a strong temperature dependence of the peak heights and positions.
(b-d) $\chi_{\rm{ac}}(H)$, its field derivative, and $\chi''(H)$ at $T=277.5$~K.
The inflection points in \chiach (vertical lines) are chosen to define the various phase boundaries.
Three of them, which are supposedly indicating first-order transitions, coincide with distinct maxima in $\chi''(H)$.
\label{fig:chiHrun}}
\end{figure}

\begin{figure}
\begin{center}
\includegraphics[width=1\textwidth]{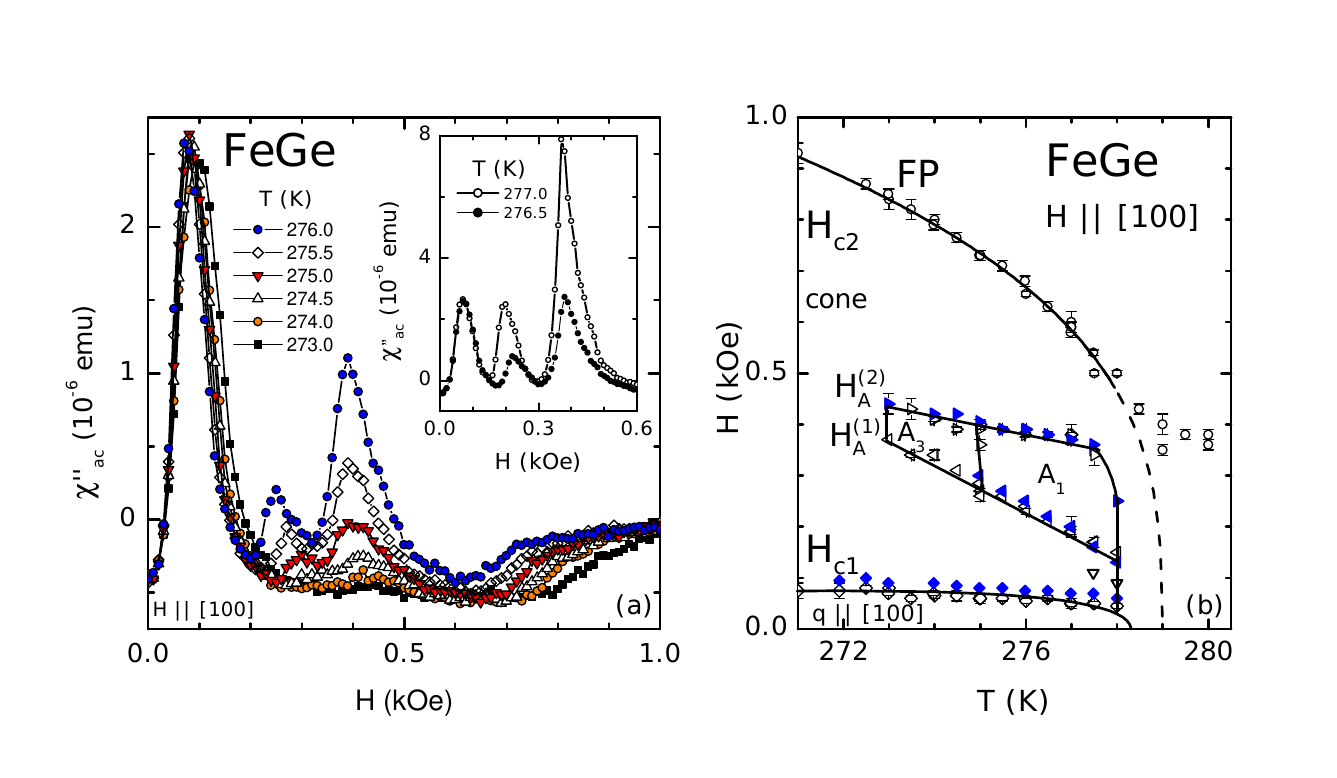}
\vspace*{-10mm}\caption{(a) Field-induced variation of $\chi''_{\rm{ac}}$, the imaginary part of $\chi_{\rm{ac}}$,
at various temperatures.
Apart from the first pronounced peak, which is related to $H_{c1}$, the strength of the other peaks changes strongly with decreasing temperature.
They are attributed to the field boundaries $H_A^{(1)}$ and $H_A^{(2)}$.
(b) Phase diagram of FeGe based on the \chiach data.
$H_{c1}$ delimits the helical from the cone phase and at $H_{c2}$ the field-polarized (FP) state is reached.
The $A$ region is indicated by the open triangles.
The peak positions in $\chi_{\rm{ac}}''(H)$ (bold symbols) agree very well with $H_{c1}$ and the $A$-phase
boundaries, $H_A^{(1)}$ and $H_A^{(2)}$, albeit no signature of $H_A^{(1)}$ was observed in $\chi''_{\rm{ac}}$ below 275~K.
The extrapolation of $H_{c2}(T)$ (dashed line) to zero field clearly deviates from $T_C$ where helical order sets in.
\label{fig:imag}}
\end{center}
\end{figure}

\section{Results}\label{sec:results}
Figure \ref{fig:chiHrun}(a) shows the field dependence of \chiac of FeGe for temperatures below $T_C=278.2(3)$~K \cite{Wilhelm11}.
Common to all curves is a minimum in \chiac at fields of the order of a few hundred Oe.
This is the fingerprint of the $A$ phase, observed first in the isostructural MnSi by ultrasound attenuation \cite{Kadowaki82} and subsequently by \chiac \cite{Thessieu97}.
It is the phase space where theoretical considerations predicted a new (vortex) state between the uniformly magnetized high-field and the one-dimensionally helically modulated low-field state \cite{Bogdanov89,Bogdanov94}.
Well below $T_C$ and above $H_{c1}$, i.e., in the conical phase, the magnetization changes almost linearly with field and will be eventually constant above $H_{c2}$, when the field-polarized state is reached.
This means, that \chiach ($\equiv\partial M/\partial H$) is finite and only weakly field dependent
in the conical phase and approaches zero above $H_{c2}$ (corresponding to $H_{\mathrm{cone}}$ in figure \ref{phd1}(a)).
However, the reduction of \emph{M(H)} across the $A$ phase ($\partial M/\partial H<0$) due to a reconstruction of the magnetic modulations \cite{Bogdanov94} will lead to the evolution of a minimum in $\chi_{\rm{ac}}(H)$.
Thus, determining the fields and temperatures where this minimum in \chiac appears allows us to locate the $A$ phase in the ($H,T$) phase diagram.
We define the lower and upper field boundaries of the $A$ phase as $H_A^{(1)}$ and $H_A^{(2)}$, respectively, by the inflection points in $\chi_{\rm{ac}}(H)$.
This criterion gives also the lower ($H_{c1}$) and upper critical field ($H_{c2}$).
The corresponding peaks $\partial \chi/\partial H$ are clearly resolved (figure \ref{fig:chiHrun}(c)).
It turns out that the location of the inflection points, apart from the one related to $H_{c2}$, coincides with peaks in $\chi''_{\rm{ac}}(H)$ (figure \ref{fig:chiHrun}(d)).
The occurrence of peaks in $\chi''_{\rm{ac}}(H)$ indicates the first-order character of these phase transitions, already concluded from a small hysteresis observed in a field cycle made at 276~K \cite{Wilhelm11}.

The temperature variation of $\chi''_{\rm{ac}}(H)$ contains additional information relevant for the construction of the magnetic phase diagram.
Already from the data in the immediate vicinity of $T_C$ shown in the inset to figure \ref{fig:chiHrun}(a) it becomes evident that $\chi''_{\rm{ac}}(H)$ is very sensitive to subtle changes of the magnetic ordering.
At 278.5~K, very close, but slightly above $T_C$, $\chi''_{\rm{ac}}(H)$ shows no distinct features.
It is almost field independent up to 400~Oe, then smoothly increases and eventually becomes constant for $H\gtrsim 1.2$~kOe.
It is worthwhile to mention that this kind of field dependence was observed up to about 283~K.
The absence of any field dependence at higher temperatures might indicate that eventually the paramagnetic state is reached.
However, just 0.2~K below $T_C$, at 278~K, three well distinct peaks appeared in $\chi''_{\rm{ac}}(H)$.
The height of the peak related to $H_A^{(2)}$ is only a fraction of the peak corresponding to $H_A^{(1)}$.
This height ratio changes drastically within 0.5~K.
At 277.5~K both peaks are well pronounced although the height and position of the peak attributed to $H_{c1}$ is unchanged.
Moreover, the position of the peak related to $H_A^{(2)}$ shifted further up in field than the position of the peak attributed to $H_A^{(1)}$.

The influence of temperature on the peak positions and heights is shown in figure \ref{fig:imag}(a) where $\chi''_{\rm{ac}}(H)$ for temperatures across the $A$-phase region is plotted.
The height of the peak assigned to $H_A^{(2)}$ continues to increase and reaches a maximum at 277~K (inset to figure \ref{fig:imag}(a)).
It fades away below 273~K and only the peak related to $H_{c1}$ is present (main part of figure \ref{fig:imag}(a)).
The height of the peak related to $H_A^{(1)}$ however, attains its maximum already at 277.5~K and is hardly visible below 275~K.
This might indicate that the character of the phase transition at $H_A^{(1)}$ below 275~K has changed from first to second order.
The analysis of these field-sweep data can be summarized in the phase diagram shown in figure \ref{fig:imag}(b).
Well below $T_C$ it shows the helical, cone, and field-polarized phases already known from neutron scattering data \cite{Lebech89}.
The $A$-phase region exists in the temperature range $273\lesssim T \lesssim 278$~K and for fields  $H_A^{(1)}<H<H_A^{(2)}$ \cite{Wilhelm11}.
Also the position of the peaks in $\chi''_{\rm{ac}}(H)$ are plotted (bold symbols in figure \ref{fig:imag}(b)).
They are in very good agreement with the phase transition lines $H_{c1}$ and $H_A^{(2)}$ and in part for $H_A^{(1)}$.
The disappearance of the peak related to $H_A^{(1)}$ in $\chi''_{\rm{ac}}(H)$ below 275~K is interpreted as an additional
indication of a division of the $A$ region into the $A_1$ and $A_3$ pockets \cite{Wilhelm11}.

\begin{figure}
\begin{center}
\includegraphics[width=0.85\textwidth]{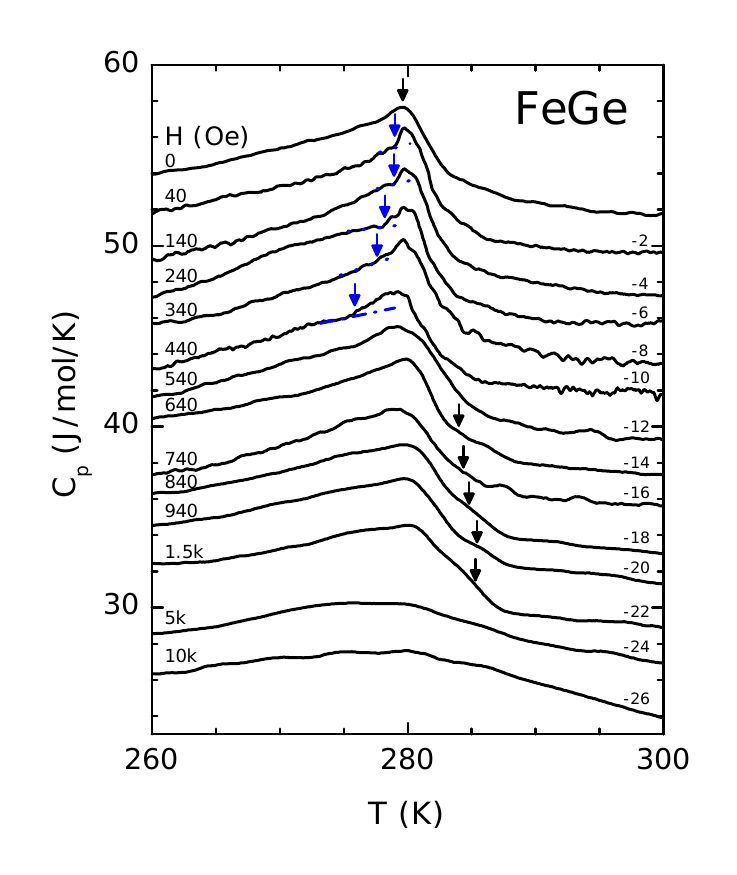}
\vspace*{-5mm}\caption{Specific heat of FeGe in the vicinity of the magnetic transition in external magnetic fields.
Curves are shifted with respect to zero field as indicated.
The position $T_0(H)$ of the peak maximum is taken as the indication of the occurrence of magnetic ordering.
A small step-like increase in $C_p(T)$ occurs just below the maximum and for fields below 340~Oe (blue arrows).
Upon increasing the field the magnetic phase transition broadens and a distinct shoulder develops above $T_0$
for $H\geq 640$~Oe (arrows).
\label{fig:cp}}
\end{center}
\end{figure}
In order to get further information on the magnetic phase transition, we measured
the specific heat, $C_p(T)$, of FeGe at various magnetic fields in the vicinity of $T_C$.
The field was along an arbitrary direction with respect to the crystallographic axes of the crystal.
Figure \ref{fig:cp} shows $C_p(T)$ curves at selected magnetic fields.
The phase transition into a magnetically ordered state in zero field is seen as pronounced maximum at $T_0=279.6$~K.
This temperature is about 1.5~K above $T_C$ where the long-period spiral propagating along equivalent [100] directions is found in polarized small-angle neutron scattering \cite{Lebech89} and where our \chiact data show a well defined maximum \cite{Wilhelm11}.
The maximum in the zero-field $C_p(T)$ data signals that some kind of magnetic order already establishes at $T_0$, i.e., prior to the long-ranged helical order setting in at $T_C$.
The shape of the maximum resembles a slightly broadened $\lambda$-anomaly, characteristic for a magnetic transition with fluctuations.
A distinct shoulder becomes clearly separated from the maximum beyond 340~Oe (indicated by arrows in figure \ref{fig:cp}).
The broadening of the maximum in field indicates that the growth of the correlation length upon approaching magnetic ordering within the critical fluctuation regime is suppressed by some kind of magnetic length scale, which drives the transition more and more towards a continuous crossover.
This is analogous to a textbook 3D-Heisenberg ferromagnetic transition but some intrinsic broadening remains down to the lowest fields suggesting the presence of additional finite-size effects, associated with a length scale
similar to effects of a domain texture.
Only a very broad magnetic contribution to the specific heat remains above 5~kOe.
This is the usual signature of the crossover from a paramagnetic to a field-polarized state.
We used the maximum in $C_p(T)$ to extract the onset of magnetic ordering and the inflection point (pronounced minimum in the derivative of $C_p(T)$) above the maximum to locate the shoulder.
Note, that some structure in form of a small upward step may be present just below the $C_p(T)$ maximum at low fields (indicated by arrows in figure \ref{fig:cp}).
This step is nearly hidden in the noise but occurs reproducibly in repeated runs in the same applied field.
High-resolution AC specific-heat measurements are needed to probe this structure in more detail.

\begin{figure}
\begin{center}
\includegraphics[width=0.85\textwidth]{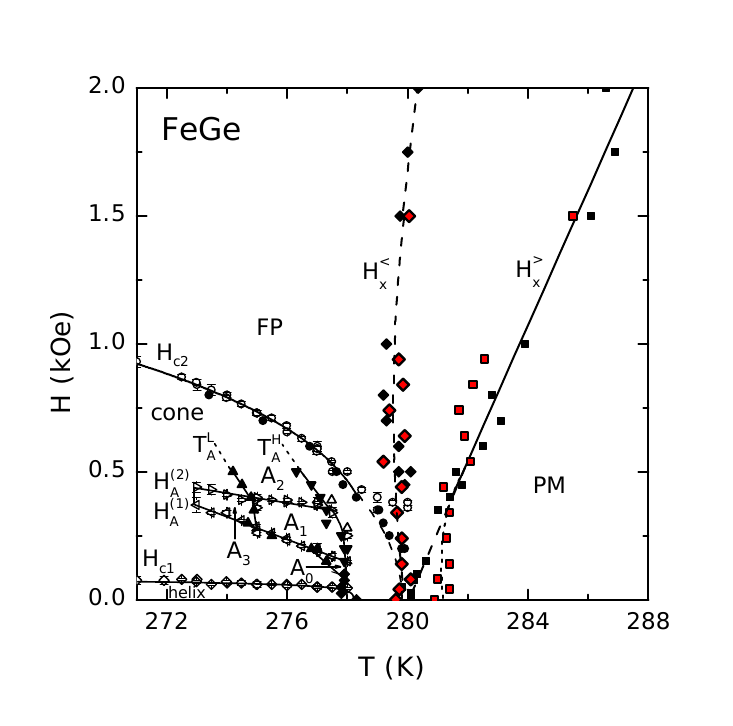}
\caption{$(H,T)$-phase diagram of FeGe based on ac-susceptibility and specific heat data.
Open and bold symbols represent data obtained from \chiach and $\chi_{\rm{ac}}(T)$, respectively \cite{Wilhelm11}.
The position of the maximum in $C_p(T)$ and the location of the inflection point at higher temperature are shown as colored symbols along the $H_{\times}^{<}(T)$ and $H_{\times}^{>}(T)$ lines, respectively.
The zero-field $C_p(T)$ data revealed some kind of magnetic order setting in at $T_0=279.6$~K.
This second order transition is distinct from the helical order found below $T_C=278.2$~K.
The phases below $T_C$ are (i) a helical state with $\bi{k}\parallel [100]$ ($H<H_{c1}$), (ii) a conical helix
phase (cone), (iii) a field-polarized state (FP, $H>H_{c2}$), and (iv) a complex $A$ region with several pockets.
The lines $H_{\times}^{<}(T)$ and $H_{\times}^{>}(T)$ indicate a crossover from the paramagnetic phase (PM)
to the FP state. \label{fig:combinedPhaseDia}}
\end{center}
\end{figure}

The field dependence of these two temperatures is added to the phase diagram of FeGe presented in figure \ref{fig:combinedPhaseDia}.
This diagram includes the field-sweep data of figure \ref{fig:imag}(b) as well as data obtained from $\chi_{\rm{ac}}(T)$ \cite{Wilhelm11}.
Below $T_C=278.2$~K and zero field the helical modulation propagates along equivalent [100] directions.
An external magnetic field of the strength of the lower critical field $H_{c1}$ forces the propagation direction along the field.
Above $H_{c1}$ a longitudinal cone is formed with the cone opening angle closing at $H_{c2}$ where the field-polarized state is reached.
These phases have been found in FeGe \cite{Lebech89}, MnSi \cite{Ishikawa76} and in pseudo binary systems
Fe$_{1-x}$Co$_x$Si \cite{Beille81,Beille83,Grigoriev07}, Mn$_{1-y}$Co$_y$Si \cite{Beille83,Bauer10},
and Mn$_{1-z}$Fe$_z$Si \cite{Bauer10,Grigoriev09}.
The $A$ region found in FeGe is to some extend peculiar as it shows subtleties not observed for other
cubic helimagets.
In FeGe the $A$ region is split into several pockets \cite{Wilhelm11}.
The prominent part, labeled $A_1$, is separated by phase transition lines $H_A^{(1)}$ and $H_A^{(2)}$ from the conical phase and the $A_2$ pocket, respectively.
The latter has no clear upper field boundary and seems to transform smoothly into the conical state.
Distinct from these pockets and the conical phase two other pockets, $A_0$ and $A_3$, can be identified.
The existence of the two crossover lines $H_{\times}^>$ and $H_{\times}^<$ found in $\chi_{\rm{ac}}(T)$ \cite{Wilhelm11} is now confirmed by the specific heat data (colored symbols in figure \ref{fig:combinedPhaseDia}).
This shows that a simple picture of a metamagnetic-like transition between paramagnetic and field-polarized
state like in MnSi \cite{Yamada04} and other 3d and 5f itinerant systems \cite{Goto01} is not applicable.
In this crossover region ($T>T_C$ and $H>0$) complex field- and temperature-induced magnetic reorientation
processes take place which might lead to the formation of short-ranged chiral modulations.
Such precursor phenomena have been predicted for cubic helimagnets \cite{Roessler06,Roessler11,Leonov10}.

\section{Discussion}\label{sec:discussion}
\begin{figure}
\begin{center}
\includegraphics[width=0.9 \textwidth]{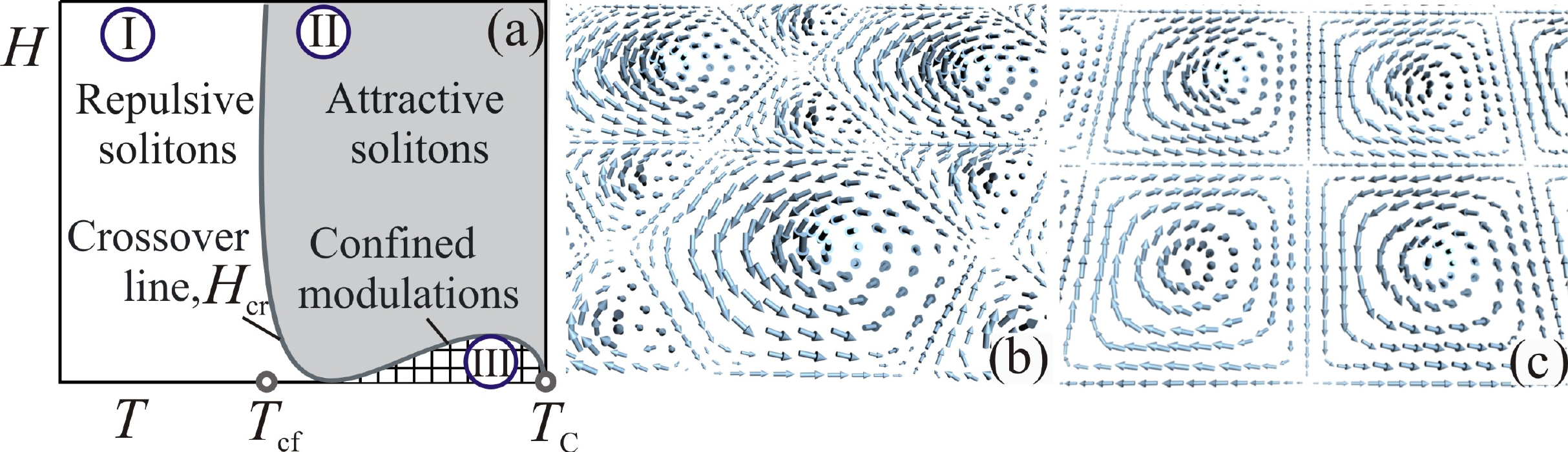}
\caption{%
(a) Calculated phase diagram for cubic helimagnets. Near the ordering temperature the critical line $H_{\mathrm{cr}} (T)$
(\ref{criticalline1}) separates the main part of the phase diagram (I) from regions with attractive kinks
and Skyrmions (II) and from confined modulations (III).
(b) Fragment of a hexagonal and (c) a square Skyrmion lattice which exist in the confinement pocket.
\label{hexsquare}
}
\end{center}
\end{figure}

The results of the previous section demonstrate a remarkable complexity of the magnetic
phase diagram of FeGe, both below and above the Curie temperature.
In combination with experimental results on MnSi \cite{Pappas09,Pappas11},
Mn$_{1-y}$Co$_y$Si \cite{Grigoriev09}, and Mn$_{1-z}$Fe$_z$Si \cite{Grigoriev11}
obtained by different neutron scattering techniques this provides profound evidence for
complex magnetic states in the precursor region of cubic helimagnets.
Theoretically these phenomena can be understood by the nucleation
of modulated magnetic states composed of confined chiral solitons as described in section \ref{sec:basicTheory}.
For the Dzyaloshinskii model of chiral magnets (\ref{density}), it is found that these solitons are coupled through
attractive inter-soliton interactions above the confinement temperature, as defined in (\ref{exchangeshift}).
More precisely, the interaction energy between isolated kinks or between Skyrmions shows damped oscillations above the confinement temperature, and the inter-soliton potential develops negative attractive wells
for certain distances between the soliton cores.
Therefore, bound soliton states can be formed \cite{Wilhelm11,Roessler11,Leonov10}.
Thus, this confinement temperature, which is near the ordering temperature, marks a crossover
where the nucleation transition should evolve differently than at lower temperature.
A detailed analysis of Skyrmion solutions for the isotropic model $w= w_0 (\bi{M}) + f(M)$ (\ref{density})
shows \cite{Roessler11,Leonov10} that the region with attractive Skyrmions in the phase diagram (figure \ref{phd1}(a))
is separated from the regular region by a crossover line \cite{Wilhelm11,Leonov10}
\begin{eqnarray}
H_{\mathrm{cr}} =H_0 \sqrt{(1 \pm \nu)} (1/2 \pm \nu )^2, \quad
T = T_{\mathrm{cf}} + \Delta_D \nu^2,
\label{criticalline1}
\end{eqnarray}
with $H_0 = \Delta_D \sqrt{2 \Delta_D /b}$, the running parameter $\nu$, and $T_{\mathrm{cf}}$, the confinement
temperature at zero field.
This crossover line is shown in the calculated phase diagram presented in figure \ref{hexsquare}(a).
Importantly, similar effects occur for 1D chiral solitons (kinks),
where the crossover line for kinks and Skyrmions coincide \cite{Schaub85,Yamashita85}.
This crossover of inter-solitonic interactions is caused by the "softening" of the magnetization modulus and the coupling
between angular (twisting) and longitudinal modulations of the magnetization.
Correspondingly, rigorous numerical solutions for helicoids and Skyrmion lattices in this region show that magnetic
states are marked by a strongly inhomogeneous magnitude of the magnetization modulus \cite{Roessler06,Roessler11,Leonov10,LeonovThesis}.

Thus, the confinement temperature $T_{\mathrm{cf}}$
separates the region of the ordered (modulated) states into two different parts.
In the main part of the magnetic phase diagram ($T < T_{\mathrm{cf}}$) skyrmionic and helical textures consist of
localized \textit{repulsive} cores (figure \ref{phd1} and \ref{hexsquare}).
The shape of these \textit{regular} chiral modulations is determined by the
strong exchange and DM interactions.
These localized modulations are stable or meta-stable against any weaker internal and external perturbations.
However, above the confinement temperature ($T_{\mathrm{cf}}< T < T_{\mathrm{C}}$),
the oscillating and competing character of
the inter-soliton couplings strongly influences the formation of the modulated states and their evolution in applied fields \cite{Leonov10}.
In particular, they establish a \textit{confinement} pocket (figure \ref{hexsquare}(a)) where
chiral modulations exist only as bound states, i.e., hexagonal and square Skyrmion lattices (figure \ref{hexsquare}(b,c)).

The nucleation transition to form modulated magnetic states requires the simultaneous nucleation and condensation
of solitonic units.
Detailed numerical studies for the isotropic Dzyaloshinskii model in this region \cite{LeonovThesis}
show that, owing to the attractive coupling between these stable units, clusters or extended bound configurations
(mesophases) have lower energy than the isolated solitons.
Different configurations of such clusters, however, are separated by tiny energy barriers.
Such characteristic features of confined modulations underlie the numerous magnetic peculiarities reported
near the ordering temperature of the cubic helimagnets:
$T_{\mathrm{D}}$ extends above the Curie temperature $T_{\mathrm{C}}$.
Therefore, magnetic precursor fluctuations are perceived well above
establishing true magnetic long-range order in the form of a helicoidal or conical helix state.
This is the region of distinct anomalies between the paramagnetic state and the
helical states at zero field, and the paramagnetic region and the field-polarized state, i.e.,
between the fields $H_x^{<}$ and $H_x^{>}$ (see figure \ref{fig:combinedPhaseDia}).
As the formation of a chiral condensate, composed from these localized
entities, is not accompanied by a diverging correlation length, only a smooth
crossover is expected.

\begin{figure}
\begin{center}
\includegraphics[width= 0.7 \textwidth]{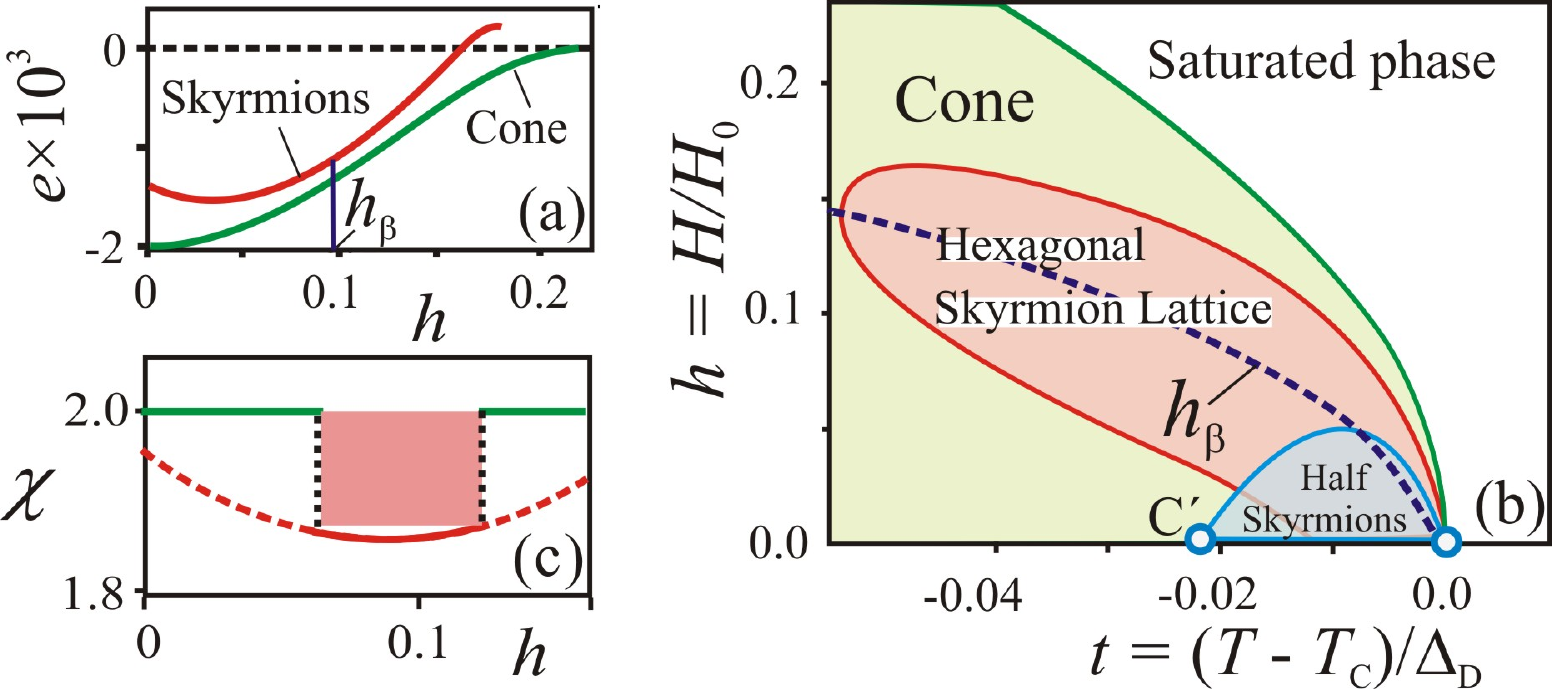}
\caption{%
(a) The energy difference $\Delta e$ between the equilibrium energy of the cone and saturated state (green line) and the hexagonal Skyrmion
lattice and the saturated state (red line) as function of the magnetic field $h= H/H_0$ for $t = (T-T_{\mathrm{C}})/\Delta_{\mathrm{D}}=-0.02$.
At a field $h_{\beta}$ the energy difference between these phases
reaches a minimum.
The dashed line $h_{\beta} (t)$ in (b) shows the temperature
dependence of these points in the phase diagram.
(b) The magnetic phase diagram of cubic helimagnets
for applied fields along the [001] axes.
For an exchange anisotropy $K/A = -0.05$ there is a region with a thermodynamically stable hexagonal Skyrmion
lattice.
A half Skyrmion square lattice corresponds to the ground
state of the systems for $T_{\mathrm{C'}} < T < T_{\mathrm{C}}$.
(c) The numerically calculated magnetic susceptibility for $t = -0.02$ indicates
the transition into the Skyrmion states (see figure \ref{fig:chiHrun}).}
\label{phd2}
\end{center}
\end{figure}

A detailed analysis of rigorous solutions for the isotropic model $w_0 (\bi{M}) + f(M)$ (\ref{density})
shows that for this functional the cone phase has the lowest energy compared to other modulated phases
within all regions where modulated states exist \cite{Leonov10}.
However, for $T_{\mathrm{cf}}< T < T_{\mathrm{C}}$ the oscillatory, competing character of the inter-soliton
couplings is confounding the hierarchy of magnetic interactions characteristic for low temperature modulations and
increases the role of minor couplings such as anisotropies, dipolar interactions, or fluctuations etc.
Small "anisotropic" energy contributions to the isotropic model can drastically change the energy balance and stabilize alternative modulated states.
To illustrate this fact we present the calculations of the equilibrium energies for the cone
and hexagonal Skyrmion lattice phases within the isotropic model (figure \ref{phd2}(a)).
Importantly, the energy difference between these two phases is minimal at a certain field
$h_{\beta}$ and gradually decreases to zero as temperature approaches the Curie temperature ($t = (T-T_{\rm{C}})/\Delta_{\rm{D}}=0$).
In the vicinity of the Curie temperature and around the line $h_{\beta} (t)$, thus, minor effects can stabilize
skyrmionic phases.

In addition, we have calculated the magnetic phase diagram for the Dzyaloshinskii model of chiral cubic helimagnets,
including cubic exchange anisotropy.
This is the classical model that has been used earlier to analyze the magnetic states
in chiral cubic helimagnets \cite{Bak80}.
The phase diagram with exchange anisotropy $K = -0.05 A$ (figure \ref{phd2}(b)) displays an extended area where
the hexagonal Skyrmion lattice corresponds to the global minimum of the system.
This area is located around the $h_{\beta}(t)$ line and has a characteristic
elongated shape similar to the \emph{A} pockets observed in cubic helimagnets (c.f. figure \ref{fig:combinedPhaseDia}).
Furthermore, these calculations also yield a stable square half-Skyrmion lattice as ground state for a temperatures $T_{\mathrm{C'}} < T < T_{\mathrm{C}}$ between the paramagnetic $ T > T_{\mathrm{C}}$ and the helical phase ($ T< T_{\mathrm{C'}})$.
The formation of this stable skyrmionic texture suggests the spontaneous formation of a further competing
type of precursor in the temperature-driven magnetic ordering process with split skyrmionic units and defect points or lines, where the magnetization passes through zero (figure \ref{hexsquare}(c)).
At lower temperatures, long-range ordered condensed phases may finally form.
The onset of the helical order occurs at $T_{\mathrm{C'}}$ i.e., below $T_{\mathrm{C}}$ owing to the competition
with spontaneous half-Skyrmion states \cite{Roessler06}.

It is worth noting that another important hint to resolve the \emph{A}-phase puzzle arises from the comparison of
the observed (e.g. figure \ref{fig:chiHrun}) and the calculated $\chi(H)$ dependence (figure \ref{phd2}(c)).
Experimentally the characteristic U-shaped anomaly in \chiach indicating the \emph{A}-phase area has also been observed in MnSi \cite{Gregory92,Thessieu97},
Mn$_{1-y}$Co$_y$Si and Mn$_{1-z}$Fe$_z$Si \cite{Bauer10}.
These observed anomalies are very similar to the calculated functions $\chi(H)$ shown in figure \ref{phd2}(c) which
correspond to the hexagonal Skyrmion lattice.

Interestingly, the calculations of the magnetic phase diagram in the framework of a modified Dzyaloshisnkii model \cite{Roessler06,Leonov10}
also indicated the half-Skyrmion lattice as the ground state in the
vicinity of the ordering temperature without anisotropic coupling
terms.
It also  yields an elongated region located around $h_{\beta} (t)$ with stable helicoids.
This model also contains a phase region for a densely packed hexagonal
Skyrmion lattice, but with a reversed topological charge of the unit cell, i.e.,
a $+\pi$-Skyrmion lattice, in contrast to the $-\pi$-Skyrmion lattices
found in the common Dzyaloshinskii-model (for details see \cite{Wilhelm11,Leonov10}).
The comparison between such models shows that details of possible phases and
phase diagrams depend very sensitively on model details in the precursor region
of chiral helimagnets.
Therefore, a resolution of the precise magnetic structure in the precursor region requires a
very detailed understanding of various additional influences in a realistic model.

\section{Conclusion}
The ac-magnetic susceptibility data of cubic FeGe yield a complex
sequence of different phase transitions and crossovers in the vicinity of
$T_{\mathrm{C}}=278.2(3)$~K for fields applied parallel to the [100] direction.
Apart from the already known helix, cone, and field-polarized states the \emph{A}-phase region was found.
The U-shape anomalies in $\chi_{\rm{ac}}$ as well as pronounced peaks in its imaginary part allowed the \emph{A}-phase boundaries to be determined.
A segmentation of the \emph{A}-phase region is concluded from the combined \chiact and \chiach runs and subtleties in $\chi_{\mathrm{ac}}''(H)$.
The field-dependence of the specific heat data confirmed the existence of a broad crossover region between the field-polarized and paramagnetic state.
They yield a $\lambda$-like anomaly at $T_0=279.6$~K (at $H=0$), which is typical for a magnetic transition accompanied by fluctuations.
The field-induced broadening of this anomaly and its shift in temperature indicates that the transition evolves towards a continuous crossover.
The usual crossover from the paramagnetic to field-polarized state is observed above 5~kOe.
These data show that FeGe provides a generic example of a chiral cubic helimagnet
with a complex magnetic phase diagram in a narrow precursor region.
It extends about 7~K below $T_0 \simeq 280$~K where some kind of magnetic precursor fluctuations are perceived.
Furthermore, it was possible to obtain an estimate for the confinement temperature $T_{\rm{cf}}\approx 273$~K as well as for the exchange shift $\Delta_D\approx 6$~K of the ordering temperature.

The theoretical interpretation of these results is based on the change of the character of inter-soliton interactions
from being attractive near the Curie temperature to repulsive in the remaining part of the magnetic phase diagram.
A critical phase line, $H_{\mathrm{cr}}(T)$, separates these two parts and defines a confinement region where
specific chiral modulations exist.
This crossover of the inter-core coupling and the onset of confined modulations are two fundamental phenomena.
They provide the first realistic mechanism underlying the rich physics of "precursor effects" in chiral helimagnets.
The experimental and theoretical findings require a substantial revision of the existing picture of
the ordering transitions in cubic helimagnets.
In a new paradigm of \textit{confined chiral precursor states},
the heuristically introduced "precursor region" must now be understood by
(i) the existence of magnetic localized textures with attractive core-core interactions,
(ii) a competition between different types of chiral modulations, i.e., kinks and Skyrmions, and
(iii) the formation of dense assemblies of these solitonic units.
The inherent frustration in the chiral coupling mechanisms then can lead to the formation of various \textit{mesophases} in very narrow temperature-field ranges.
Importantly, the crossover of the inter-core coupling and the formation
of specific precursor modulations is a general phenomenon attributed to \textit{all} chiral systems,
including non-centrosymmetric magnets, chiral liquid crystals, and multiferroics.

\ack
We acknowledge fruitful discussions with C. Pappas and Yu. Grin as well as
technical assistance from R. Koban, Yu. Prots, and H. Rave. Support by DFG project RO 2238/9-1 is
gratefully acknowledged.

\end{document}